\documentstyle[12pt]{article}
\columnwidth\textwidth
\newcommand{\vfacs}{\sqrt{2m_{B_1}} \, \int
            \frac{d^3k_1}{\sqrt{4 k^0_1 k^0_3} \, (2\pi)^3} \,
\psi_1      ( \vec{k}_1        )  }

\newcommand{\vac}{\left|0\right>}

\newcommand{\FF}{F(\vec{q}\,\vphantom{\vec q}^2)}
\newcommand{\mba}{m_{B_1}}
\newcommand{\mbb}{m_{B_2}}
\newcommand{\qsl}{q \hspace{-6pt} / \hspace{4pt}}

\def\pushq{\vphantom{\vec q}\,}
\begin{document}
\thispagestyle{empty}   
\begin{flushright}
SLAC--PUB--6487      \\
ZU-TH 5/94\\
April 1994 \\
(T/E)
\end{flushright}
\begin{center}
  \begin{Large}
  \begin{bf}
  Atomic Alchemy
\footnote{Work partially supported by Schweizerischer
Nationalfonds and the Department of Energy under contract
DE--AC03--76SF00515.}
  \end{bf}
  \end{Large}
\end{center}
 \vspace{.7cm}   
\begin{center}
  \begin{large}
    C. Greub and D. Wyler \\
    \vspace{0.5cm}
    Institut f\"ur Theoretische Physik,
    Universit\"at Z\"urich \\ Z\"urich, Switzerland  \\
  \vspace{0.8cm}
    and  \\
  \vspace{0.8cm}
    S. J. Brodsky and C. T. Munger   \\
    \vspace{0.5cm}
    Stanford Linear Accelerator Center \\
Stanford University, Stanford, California 94309 \\
  \end{large}
\vspace{0.8cm}
\end{center}
\begin{abstract}
We consider the transitions between electromagnetic bound states,
such as the exclusive weak decay
of a muonic atom into an electronic
atom: $(\mu^- Z)  \to (e^- Z) {\bar \nu_e }\nu_\mu . $  We show
that relativistic effects in the atomic
wavefunctions are crucial for determining
the rate.  In the case of heavy atoms, the
exclusive channel branching ratios exceed  $10^{-6},$
possibly bringing
the study of these rare decays within experimental reach.
Such processes thus provide a detailed laboratory
for studying the high momentum tail of  wavefunctions
in atomic physics; in addition, they provide a simple
toy model for investigating analogous exclusive heavy
hadronic decays in quantum chromodynamics  such as $B \to \pi e \nu.$
\vspace{0.8cm}
\begin{center}
(Submitted to Physical Review D.)
\end{center}
\vspace{0.8cm}
\end{abstract}

\newpage
{\bf 1.)  Introduction}
\vspace*{.8cm}

The calculation of  the rate of a weak decay of a heavy hadron
into an exclusive channel, such as $B \to D e \nu$ or
$B \to \gamma K^*$,  poses a challenging
problem in nonperturbative
quantum chromodynamics because
all the complexities of the hadronic wavefunctions
enter. In this paper we point out the possibility
of studying an analogous, but far simpler
process in atomic physics: the weak decay of
one electromagnetic bound state into another.  Simple examples occur
in the decay of muonic atoms, such as
\begin{eqnarray}
\label{eq1}
(\mu^- Z)  \to (e^- Z) {\bar \nu_e }\nu_\mu \ ,\quad
\nonumber \\
(\mu^+ e^-) \to  (e^+ e^-)  \nu_e  {\bar \nu_\mu} \ .\quad
\end{eqnarray}
Because the atom remains in a bound state, but the
atomic or nuclear species is changed, we refer to
such processes as ``atomic alchemy."
We could also consider the decay
$(\pi^+ e^-) \to (\mu^+ e^-) \nu_\mu$, or
the decays of  Coulomb-bound
hadronic atoms such as
$(\pi^- Z) \to (\mu^- Z) {\bar \nu_\mu}$
or $(\Sigma^- Z) \to (\bar p Z) \pi^0$;
however, the observation
of the latter transitions appears to be
impractical because of the high rate
at which hadrons are absorbed by the nucleus.

Other interesting examples of atomic alchemy occur
when a nucleus is forced to change
its state because of external scattering, while an atomic
electron remains bound.  The scattering may be due to
neutron-induced fission, nuclear Compton scattering, or
photodisintegration; the latter is particularly interesting
for deuterium,
$\gamma (d^+ e^-) \to (p^+ e^-) n$.
One can also study atomic alchemy when a nucleus changes its
state by an internal process, such as $\beta$-decay.

In this paper, we consider the exclusive
transition $B_1 \to B_2 + X$, where both
$B_1$ and $B_2$ are bound states consisting of particles $(b_1 s)$
and
$(b_2 s)$, respectively,
and the transition proceeds by the
weak decay $b_1 \to b_2 + X$.  The particle
$s$ will be called the spectator.
We assume that there is an empty bound state in $B_2$ for the particle
$b_2$ to settle in;
in the detailed calculations given below we take
this to be the ground state.

The basic features of atomic alchemy
can  be easily understood using non-relativistic mechanics; the analysis
is similar to that for nuclear collisions involving capture of atomic
electrons, as discussed, for example, by Migdal \cite{Migdal}.
We review the major points before plunging into the
full relativistic calculation.
Because the weak decay
occurs over a time short compared to the period of an orbit,
we can use the  ``sudden approximation''.
The probability amplitude for the atomic transition then
factors into the free matrix element of the  weakly
decaying (moving) particle times a form factor $\FF$.
This form factor is just
the overlap
of the wavefunctions of the initial and final states, which we write
as $\psi_1$ and $\psi_2$.
In the rest frame of
the initial state, $\FF$ is
\begin{equation}
\label{eq2}
\FF =
\int \frac{d^3 k_1}{(2\pi)^3} \,
\psi_{1}(\vec{k_1}) \,
\psi^*_{2}(\vec{k_2}=m_{red,2} \,\vec v_{rel,2})
\quad.
\end{equation}
Here $\vec k_1$ is the momentum of $b_1$
(see Fig.~1), and
$m_{red,2}$ and     $\vec v_{rel,2}$
are the reduced mass and the relative velocity of the final
state particles $b_2$ and
$s$.  The velocity $\vec v_{rel,2}$
is a function of
$\vec q$, the momentum transfer carried by $X$,
which by conservation of momentum is equal to the recoil momentum
of $B_2$.  We have
\begin{equation}
\label{vrel}
\vec{v}_{rel,2} = \frac{\vec{k}_1 - \vec{q}}{m_{b_2}} +
\frac{\vec{k}_1}{m_s}     =
\frac{\vec{k}_1}{m_{red,2}} - \frac{\vec{q}}{m_{b_2}} \quad .
\end{equation}

In the decay $(Z \mu^-) \to (Z e^-) \nu_\mu
\bar{\nu_e}$, the argument
$\vec{k}_2$ of the wavefunction $\psi_2^*$
is approximately  $\vec{k}_2\approx\vec{k}_1
- \vec{q}$.
In momentum space
the muon wave function does not vary much over the domain
where the electron wave function differs appreciably from zero.
The integral in
Eq.~(\ref{eq2}) can therefore be approximated by
\begin{equation}
\label{eq3}
\FF =
\psi_{1}(\vec{q\,}) \,
\psi^*_{2}(\vec{r}=0)
\quad.
\end{equation}
For small $Z$ the momenta of both the muon and the electron are
nonrelativistic
and the effects of the finite nucleus size are small;
we therefore can use for $\psi_1$ and $\psi_2$
the Schr\"odinger
wavefunctions for a point nucleus.
For the $1S$ states
\begin{equation}
\label{schroed}
\psi_i(\vec{k}) = \frac{8 \sqrt{\pi} a_i^{3/2}}{
\left[ 1 + a_i^2 \, \vec{k}^2
\right]^2} \ , \quad a_i =
\frac{1}{m_{red,i} Z \alpha} \ , \ (i=1,2) \ ,
\end{equation}
where
$a_i$ denote the Bohr-radii of the atoms.
Because $\psi_i(\vec{r}=0) = [\sqrt{\pi} a_i^{(3/2)}]^{-1}$,
Eq.~(\ref{eq3}) becomes
\begin{equation}
\label{festimate}
\FF = 8 \, \biggr({m_e \over m_\mu}\biggr)^{3/2}
\left[{1 \over  1+ \vec{q \,}^2/ (Z \alpha m_\mu)^2
}\right]^2 \quad .
\end{equation}

The physics of capture into a hydrogenic $nS$ state
is controlled by the ratio
of the change in nuclear velocity to the atomic velocity
$Z \alpha /n.$
Equation~(\ref{festimate})
suggests that the highest rates for the transition of
muonic
to electronic atoms
will be found
for heavy nuclei with $(Z \alpha m_\mu)^2 \ge \vec{q\,}^2$.
For these atoms,
however,
Eq.~(\ref{festimate}) provides only a very rough
estimate for the form factor because
relativistic effects (mainly affecting the wavefunction of
the $(Z e^-)$ atom) become crucial and
greatly enhance the transition rate.

In the case of the transition
$(\pi^+ e^-) \to (\mu^+ e^-) \nu_\mu$ the monoenergetic weak decay
changes the species and velocity of the heavy particle serving
as a nucleus.
Since the momentum transfer from the pion
to the muon is much larger than either Bohr
momenta,
the integrand in the
form factor in Eq.~(\ref{eq2})
is dominated by those values
of $k_1$ for which either $k_1$ or $k_2$ matches a Bohr momentum.
Both contributions turn out to be
equally important, and
using for the $1S$ states
the Schr\"odinger
wavefunctions for a Coulomb potential, we obtain for the form factor
\begin{equation}
\label{eq3a}
\FF \simeq 2
\psi_{1}(\vec{r}=0)
\psi^*_{2}(m_{red,2} \vec{v}_{rel,2})  \simeq
16 \left( \frac{\alpha}{v_{rel,2}} \right)^4
\quad.
\end{equation}
Here $v_{rel,2} =
(m^2_\pi-m^2_\mu)/(2 m_\pi m_\mu) = 0.28,$
which is small enough
that our non-relativistic treatment is justified.
The probability for capture  of the electron from the muonic decay
is unfortunately very small:
\begin{equation}
\label{eq4}
P = [\FF]^2 \simeq
256 \alpha^8/v_{rel,2}^8 \sim 5 \times 10^{-11} \quad .
\end{equation}

\vspace*{1cm}
{\bf 2.) Relativistic Analysis of Atomic Transitions}
\vspace*{.8cm}

We now turn to a more detailed analysis
which allows us to treat decays such as
$(Z \mu^-) \to (Z e^-) \nu_\mu
\bar{\nu_e} $, where
for large $Z$ the velocities of the bound constituents
are relativistic. In principle, the analysis requires a
fully covariant description, such as the Bethe-Salpeter equation
or the light-cone Fock state expansion. In Ref.~\cite{GW}
a  covariant description was given, but at the expense of a
somewhat artificial momentum-dependent mass for the decaying
constituent. While that description
works well when
the decaying constituent is much heavier than the spectator,
it fails when the spectator is the heavier particle,
because the high momenta of the constituents, which prove crucial,
are cut off
(see Eq.~(2.1) of Ref.~\cite{GW}).
Here we use a non-covariant description that is adequate for the
cases of interest, where the spectators in the initial and final
bound states have small relative velocity, but where the velocities of
the particles bound to the spectators
can be relativistic.  This  description is particularly appropriate
for the transition
$(\mu^- Z) \to (e^- Z) + \nu_\mu \bar{\nu}_e$
because the mass of a nucleus is much greater than $m_\mu$.

We begin with a relativistic treatment of
the simple, monoenergetic, two-body atomic transition
$(\pi^+ e^-) \to (\mu^+ e^-) + \nu_\mu$.
This is a close analog to the hadronic decay $B \to D e \nu.$
The relative velocity of the $(\pi e)$ and $(\mu e)$ systems
is only $v_{rel} = 0.28$, so one can simplify the kinematics
by assuming them to be at relative rest.
We consider only $S$-wave bound states, so the spin of the
bound state $B_1$ is just the spin of the electron, and the final
bound state $B_2$ is either a pseudoscalar or a vector particle
with the corresponding
well-known spin combinations of the constituents.
For example,
the initial state $B_1$ with spin projection $R$
is represented in its rest frame as
\begin{eqnarray}
\label{bstate}
\left|B_1,\vec{p}_B=0,R\right> &=& \vfacs \,
a^+_R(\vec{k}_1) \, b^+(\vec{k}_3) \vac \quad ,
\end{eqnarray}
where $\vec{k}_3=-\vec{k}_1$, and
$k_i^0=(\vec{k}_i^2+m_i^2)^{1/2}$ for ($i=1,3$), and
$a^+(\vec{k}_1)$ and
$b^+(\vec{k}_3)$ are creation operators for the constituents that
act on the vacuum state $\left|0\right>$.
The state is ``covariantly" normalized
in the volume $V$, so that
\begin{equation}
\label{norm}
\left<B_1,\vec{p}_B=0,R|B_1,\vec{p}_B=0,S\right>= 2 m_{B_1} V \,
\delta_{RS} \quad , \, \mbox{if} \quad
\int \frac{d^3k_1}{(2 \pi)^3} |\psi_1(\vec{k_1})|^2 = 1 \quad .
\end{equation}
The matrix element for the decay
$(\pi^+ e^-) \to (\mu^+ e^-) + \nu_\mu$
is then written as
\footnote{We use spinors normalized
as $\bar{u}_r(\vec{p\,}) \, u_s(\vec{p\,}) =
  - \bar{v}_r(\vec{p\,}) \, v_s(\vec{p\,}) = 2 m \delta_{rs}$}
\begin{equation}
\label{matpi1}
M_{rs} = \frac{4 G_F V_{ud}}{\sqrt{2}} \,
\sqrt{4 \mba \mbb} \,
\int \, \frac{d^3 k_1}{(2 \pi)^3} \,
\frac{\psi_1(\vec{k}_1)}{\sqrt{2 k_1^0}} \,
\frac{\psi^\star_2(\vec{k}_2)}{\sqrt{2 k_2^0}} \, S_{rs} \,
\frac{f_\pi m_\pi}{2}
\end{equation}
\begin{equation}
\label{matpi2}
S_{rs} =  \bar{u}_r(\vec{q}\,) \gamma_0
\frac{1-\gamma_5}{2} \, v_s(\vec{k}_1 - \vec{q}\,) \quad , \quad
k_2^0=(m_\mu^2+(\vec{k_1} - \vec{q}\,)^2)^{1/2} \quad ,
\end{equation}
where $f_\pi \approx 130$ MeV is the pion decay constant,
and $V_{ud}$ is the relevant CKM~matrix element. Here
$u_r(\vec{q}\,)$ and
$v_s(\vec{k}_1 - \vec{q}\,)$ are
the Dirac spinors of respectively the neutrino
with spin $r$ and the muon with spin $s$.
The argument $\vec{k}_2$ of the bound
state wavefunction $\psi^\star_2$ is given by
$\vec{k}_2=\vec{k}_1 - (m_{red,2}/m_\mu) \vec{q}.$
Note that in $S_{rs}$ we have kept only the zeroth component
of the weak current
since the pion is essentially at rest.
In the limit $\vec{k}_1 - \vec{q} \ll m_\mu$, only
the large components of the spinor need to be retained, and
$S_{rs}$
takes the simple form
\begin{eqnarray}
\label{srs}
S_{rs} &=& \frac{1}{2} \, \sqrt{2 m_\mu |\vec{q}\,|} \,
\left( \chi_r^+ ( 1 -
\frac{\vec{\sigma}\cdot \vec{q}}{|\vec{q}\,|} \, ) \,
\varepsilon \chi_s \right) \quad,
\nonumber \\
\varepsilon &=& i \sigma_2 \ , \
\chi_1^+ =(1,0) \ , \
\chi_2^+ =(0,1) \ .
\end{eqnarray}
We square the matrix element, sum over pseudoscalar and
vector final states, and average over the spin of the initial state.
Using
\begin{equation}
\label{spinsumme}
\sum_{rs} S_{rs} S_{rs}^\star = 2 m_\mu |\vec{q}\,|
\ ,
\end{equation}
the spin-averaged matrix element becomes
\begin{eqnarray}
\label{matelav}
\overline{|M|^2_\Sigma} &=& 4 G_F^2 |V_{ud}|^2 f_\pi^2 m_\pi^2
|\FF|^2 m_\mu |\vec{q}\,|  \nonumber \\
\FF &=&
\int \, \frac{d^3 k_1}{(2 \pi)^3} \,
\psi_1(\vec{k}_1) \,
\psi^\star_2(\vec{k}_2) \quad ,
\end{eqnarray}
where we have used (see Eq.~(\ref{matpi1}))
the approximation
$4 k_1^0 k_2^0\approx 4 \mba \mbb $.
The two-body kinematics fixes the momentum transfer $|\vec{q}\,|$
to be
\begin{equation}
\label{qkin}
|\vec{q}\,| = (\mba^2 - \mbb^2)/(2 \mba) \approx
(m_\pi^2 - m_\mu^2)/(2 m_\pi) \quad .
\end{equation}
The decay rate in this approximation is
\begin{equation}
\label{width}
\Gamma = \frac{m_\pi^2 - m_\mu^2}{16 \pi m_\pi^3} \,
\overline{|M|^2_\Sigma}  \quad .
\end{equation}

To get an estimate of the errors
made by using non-relativistic mechanics, we calculated the decay rate
of a free pion making the same approximations,
i.e. just taking the zeroth
component of the weak current and retaining only the large
components of the Dirac spinor of the muon.
We find
\begin{equation}
\label{gammafreeapp}
\Gamma_{free}^{approx} = \frac{G_F^2 |V_{ud}|^2 f_\pi^2
m_\pi m_\mu
(m_\pi^2 - m_\mu^2)^2}{8 \pi m_\pi^3} \quad .
\end{equation}
This differs from the rate calculated using relativistic mechanics
only by the factor
\begin{equation}
\label{gammacomp}
\Gamma_{free}^{exact}/\Gamma_{free}^{approx} =
m_\pi/m_\mu \approx 1.3 \quad ,
\end{equation}
which indicates a possible error of about 30\% .
The branching
ratio for decay to the $1S$ state, obtained
by dividing Eq.~(\ref{width}) by the
approximate form in Eq.~(\ref{gammafreeapp}), is
\begin{equation}
\label{branching}
\mbox{BR}\big((\pi^+ e^-) \to (\mu^+ e^-) \nu_\mu\big) =
|\FF|^2 \quad .
\end{equation}
Using the Schr\"odinger
wavefunctions
of Eq.~(\ref{schroed}) (with $Z=1$ of course) we have finally that
$\mbox{BR}\big((\pi^+ e^-) \to (\mu^+ e^-) \nu_\mu\big) =  4.51
\times 10^{-11}$.

\vspace*{1cm}
{\bf 3.)  Relativistic Effects in Muonic to Electronic Atom Decays}
\vspace*{.8cm}

 We now consider
$(\mu^- Z) \to (e^- Z) \nu_{\mu} \bar{\nu}_e$, using with obvious
replacements the same kinematics as in Fig.~1.
For simplicity we will take the nucleus
to be spinless.
As we now consider a 3-body decay, we first discuss the
ranges of the relevant kinematical variables.
Because the two constituents have extremely different masses
(especially if $Z$ is large), it is useful to write the
masses $\mba$ and $\mbb$ of the bound states $B_1$ and $B_2$
in the form
\begin{eqnarray}
\label{masses}
\mba &=& M + m_1 \ , \
m_1 = m_\mu - E_{bind,1} \ ; \nonumber \\
\mbb &=& M + m_2 \ , \
m_2 = m_e - E_{bind,2} \ .
\end{eqnarray}
Here $M$ is the mass of the heavy nucleus.
In the rest frame of the decaying atom, the momentum transfer
$|\vec{q}\,|$
to the neutrino-pair
is kinematically restricted
to the range
\begin{equation}
\label{qrange}
0 \le |\vec{q}\,| \le (\mba^2-\mbb^2)/(2 \mba) = (m_1 - m_2) +
O(1/M) \quad.
\end{equation}
For a given value of $|\vec{q}\,|$, the fourth component
$q^0$ as seen in the rest frame,
and the square of the four-vector, $q^2$,
are fixed:
\begin{equation}
\label{q0qq}
q^0 = (m_1 - m_2) + O(1/M) , \quad
q^2 = (m_1 - m_2)^2 - \vec{q}\,\pushq^2 .
\end{equation}
Thus $q^2$ lies in the range $[0,(m_1 - m_2)^2]$.
The momentum transfer $|\vec q\,|$
is always very small compared to
the masses of the bound states, so the bound states
can be considered to be at
relative rest; in this approximation
our formalism will still be covariant up to
$O(1/M)$ corrections.
Writing the relevant four-Fermi operator for muon decay in charge
retention form,
the matrix element for the transition reads
\begin{equation}
\label{mat3body}
M_{sr} = \frac{4 G_F}{\sqrt{2}} \,
\sqrt{4 \mba \mbb} \, N_\mu \,
S^\mu_{sr} \quad ,
\end{equation}
where
\begin{eqnarray}
\label{nmu}
N_\mu &=& \bar{u}(p_{\nu_\mu}) \, \gamma_\mu \, \frac{1-\gamma_5}{2}
\, v(p_{\nu_e}) \\
\label{smu}
S^\mu_{sr} &=&
\int \,
\frac{d^3 k_1}{(2 \pi)^3} \, \psi_1(\vec{k}_1) \,
\psi^\star_2(\vec{k}_1-\vec{q}\,) \,
\frac{\bar{u}_s(e;\vec{k}_1-\vec{q}\,)}{\sqrt{2
k_2^0}} \, \gamma^\mu \, \frac{1-\gamma_5}{2} \,
\frac{u_r(\mu;\vec{k}_1)}{\sqrt{2 k_1^0}}  \\
k_1^0 &=& \sqrt{m_\mu^2 + \vec{k}_1^2} \ , \
k_2^0  =  \sqrt{m_{e\vphantom{\mu}}^2
+ (\vec{k}_1 - \vec{q}\,)^2} \quad .
\end{eqnarray}
The muon spin $r$ and the electron spin $s$ are just the spins of
the bound states $B_1$ and $B_2$.  In general a Wigner rotation~\cite{GW}
must be used to relate spins in two different frames, but here
the rotation
is essentially unity because the relative velocity of $B_1$ and $B_2$
is small.

We seek an expression for $M_{sr}$ that is correct to zeroth order
in $1/M$.
This is found most easily by first writing the trivial
identity
\begin{equation}
\label{trivid}
S^\mu_{sr} = \delta_{ss'} \, S^\mu_{s'r'} \, \delta_{r'r} \ ,
\end{equation}
and then by rewriting the Kronecker-deltas in terms of
the spinors for the bound
states and their constituents,
\begin{eqnarray}
\label{trick}
\delta_{r'r} &=& \left( 2 \mba (k_1^0+m_\mu) \right)^{-1/2} \,
\bar{u}_{r'}(\mu;\vec{k}_1) \, u_r(B_1;\vec{0}\,) \quad ,
\nonumber \\
\delta_{ss'} &=& \left( 2 \mbb (k_2^0+m_e) \right)^{-1/2} \,
\bar{u}_{s}(B_2;\vec{p}_{B_2}) \, u_{s'}(e;\vec{k}_1-\vec{q}\,)
\ .
\end{eqnarray}
The first relation is exact, and the
second introduces errors only $O(1/M)$. Then
$S^\mu_{sr}$ can
be written as
\begin{equation}
\label{tmudef}
S^\mu_{sr} = (4 \mba \mbb)^{-1/2} \,
\bar{u}_{s}(B_2;\vec{p}_{B_2}) \, T^\mu \,  u_r(B_1;\vec{0})
\quad ,
\end{equation}
where $T^\mu$ may be derived using eqs.~(\ref{smu}),~(\ref{trivid}),
and~(\ref{trick}).
By repeatedly using the Dirac equations
\begin{equation}
\label{diraceqapp}
\gamma^0  u_r(B_1;\vec{0}\,) =  u_r(B_1;\vec{0}) \ , \quad
\bar{u}_{s}(B_2;\vec{p}_{B_2}) \, \gamma^0 =
\bar{u}_{s}(B_2;\vec{p}_{B_2}) + O(1/M)\ ,
\end{equation}
and dropping terms $\sim q^\mu$, which vanish when contracting
with the neutrino tensor $N_\mu$ of Eq.~(\ref{nmu}), we arrive
after some not completely straightforward algebra at
\begin{equation}
\label{tmufinal}
T^\mu = F_1(q^2) \gamma^\mu L + F_2(q^2) \gamma^\mu R + F_3(q^2)
\gamma^\mu \, \frac{\qsl}{m_\mu} L +
F_4(q^2) \gamma^\mu \frac{\qsl}{m_\mu} R \ .
\end{equation}
Here $L=(1-\gamma_5)/2$, and $R=(1+\gamma_5)/2$,
and the form factors $F_i(q^2)$ are given as
\begin{equation}
\label{formgeneric}
F_i(q^2) = \int \, \frac{d^3k_1}{(2\pi)^3} \, \psi_1(\vec{k}_1)
\, \psi^\star_2(\vec{k}_1 - \vec{q}\,) \,
\frac{h_i}{\sqrt{4 k_1^0 k_2^0 (k_1^0+m_\mu)(k_2^0+m_e)}}\ ,
\end{equation}
with
\begin{eqnarray}
\label{hi}
h_1 &=& (k_1^0 + m_\mu)(k_2^0+m_e) + q^0 \,
\Big[(1-C)(k_1^0+m_\mu)-C(k_2^0+m_e)\Big]  \nonumber \\
&& \quad + (B-C)(q^0)^2 - A
\nonumber \\
h_2 &=& (C-B) q^2 - 2 A \nonumber \\
h_3 &=& \left[ (1-C) (k_1^0+m_\mu) + (B-C) q^0
\right] \, m_\mu    \nonumber \\
h_4 &=& \left[C (k_2^0+m_e) - (B-C) q^0
\right] \, m_\mu    \quad .
\end{eqnarray}
The quantities $A$, $B$, and $C$ are
\begin{equation}
\label{abc}
A= \frac{\vec{q}\pushq^2 \vec{k}_1^2 -(\vec{k}_1\cdot\vec{q}\,)^2}
{2\vec{q}\pushq^2}
\ , \
B=\frac{3 (\vec{k}_1\cdot \vec{q}\,)^2 -
\vec{q}\pushq^2 \vec{k}_1^2}{2(\vec{q}\pushq^2)^2} \ , \
C=\frac{\vec{k}_1\cdot \vec{q}}{\vec{q}\pushq^2} \ .
\end{equation}
The form factors $F_i$ may seem non-covariant,
because after the integration $d^3\vec{k}_1$
the variables $q^0$ and $|\vec q\,|$ remain as
well as the square of the four-momentum, $q^2$.
But the form factors were derived in the rest frame of $B_1$, so
$q^0$ and $|\vec{q\,}|$ are functions only of $q^2$ according
to Eq.~(\ref{q0qq}).

In terms of the quantities introduced, the matrix element
can be written in the suggestive form
\begin{equation}
\label{finalmat}
M_{sr} = \frac{4 G_F}{\sqrt{2}} \,
\bar{u}(p_{\nu_\mu})  \gamma_\mu L
\, v(p_{\nu_e}) \,
\bar{u}_{s}(B_2;\vec{p}_{B_2}) \, T^\mu \,  u_r(B_1;\vec{0})
\quad .
\end{equation}
The calculation of the decay
rate, differential
in the  momentum transfer $|\vec{q}\,|$, is now standard.
Expressing the bound state masses $\mba$ and $\mbb$ in terms of
$M$,~$m_1$, and~$m_2$ as defined in Eq.~(\ref{masses}),
one obtains
\begin{eqnarray}
\label{spec}
\frac{d\Gamma}{d|\vec{q}\,|} &=&   \frac{G_F^2
|\vec{q}\,|^2}{12\pi^3} K(|\vec{q}\,|) \nonumber \\
K(|\vec{q}\,|) &=&  [q^2 + 2 (m_1 - m_2)^2] \, (F_1^2+F_2^2) +
\frac{q^2}{m_\mu^2} \, [4(m_1-m_2)^2-q^2] \, (F_3^2+F_4^2)
\nonumber \\
&& -6q^2 \, \left[F_1 F_2 + \frac{q^2}{m_\mu^2} \, F_3 F_4 +
\frac{m_1-m_2}{m_\mu} \, (F_1-F_2) \, (F_3-F_4)\right] \quad ,
\end{eqnarray}
where the $F_i$ are the form factors defined in Eq.~(\ref{formgeneric}).
Note that $m_1$ and $m_2$ enter only through their difference;
this is clear from the decomposition (\ref{masses}),
which is invariant under the change of variables
$M \to M + \lambda$ and~$m_i \to m_i - \lambda$.

The wavefunctions that enter the form factors
$F_i$ in Eq.~(\ref{formgeneric})
are
plotted
in Figs.~2 and~3 (solid line).
For comparison
we have
also drawn the wavefunctions for a point nucleus.
The finite
size of the nucleus affects the
muon $1S$ wavefunction at high~$Z$,
as shown in Fig.~2.   The finite size also affects
the ultra-relativistic momentum tail of the
electron $1S$ wavefunction,
as shown in Fig.~3.
The calculation of these wavefunctions is discussed in detail
in the appendix.  Briefly, we have approximated
the shape of a nucleus of atomic number $A$
as a homogeneously charged sphere of radius
$r_0 = 1.3 \, A^{1/3}\,\mbox{fm}$.
In position space
we solve the Dirac equation
inside and outside $r_0$; the condition that these solutions
match at $r=r_0$ determines the ground state wavefunction.
We then take its Fourier transform.

 From the kinematics (see Eq.~(\ref{qrange}))
it is clear that
the muon $1S$ wavefunction is tested to momenta the order of $m_\mu$.
As shown in Fig.~2,
relativistic effects are moderate.
However, the finite
size of the nucleus enhances
the low momentum part of the muon wavefunction, which is fortunate
because for lower momenta the electron wavefunction
is large, and so the overlap increases.
The electron $1S$ wavefunction is also tested up to
momenta the order of $m_\mu$ and so
the ultra-relativistic tail of the Dirac wavefunction
is important.  As shown in Fig.~3 the finite size of the nucleus
diminishes the tail, and so the overlap with muon wavefunction
decreases.  At high $Z$ we find that while the shape of the spectrum
in $|\vec q\,|$ is altered, the increase and the decrease in the
total rate
due
to the finite nuclear size balance remarkably, giving the same
predicted total rate for atomic alchemy as does a calculation
using a point nucleus.

Using the same approximations
we also calculate the free-electron decay
rate $\Gamma_{e,free}$ for
$(Z \mu^-)$ decays.
We write the mass of the bound state $B_1$ as
$m_{B_1}=M +\hat{\gamma} m_\mu$, so that $\hat{\gamma} \, m_\mu$
is the total energy of the muon
(see Eq.~(\ref{qrange}).  We find
\begin{equation}
\label{totalwidth}
\Gamma_{e,free} =
\Gamma^0 \, \hat{\gamma}^2 \, \langle L^{-1}\rangle \ ,
\quad  \langle L^{-1}\rangle =
\int \,
\frac{d^3k_1}{(2\pi)^3} \, |\psi_1(\vec{k}_1)|^2 \,
\frac{m_\mu}{\sqrt{k_1^2+m_\mu^2}}  \quad,
\end{equation}
where
$\Gamma^0 =
G_F^2 m_\mu^5/(192 \pi^3)$
is the
decay width of a free muon,
and
$\left<L^{-1}\right>$ can be interpreted as the
mean inverse Lorentz factor representing the slowing of the muon
decay rate due to its orbital velocity.
Numerically, we have
$\left<L^{-1}\right>=0.96$, for $Z=80$ and $A=200$; and
$\left<L^{-1}\right>=1.00$,
for $Z=10$ and $A \le 20$.
For $Z=80$ and $A=200$, numerically ${\hat\gamma}$
is $0.91$; for a point nucleus with $Z=80$ note that $\hat{\gamma}$
and
$\left<L^{-1}\right>$
are significantly smaller,
$\hat{\gamma}= (1 - (Z \alpha)^2)^{1/2} = 0.81$
and
$\left<L^{-1}\right>=0.85$.
However, for $Z \approx 80$, muon capture dominates the
free-electron decay by a factor of about 30 \cite{aadd}.
The branching ratio is now obtained by numerically
integrating the spectrum in Eq.~(\ref{spec}) and dividing by
$\Gamma_{tot}=30 \times \Gamma_{e,free}$, with $\Gamma_{e,free}$
given
in Eq.~(\ref{totalwidth}).
Using Dirac wavefunctions,
the results for $Z=80$ are
\begin{eqnarray}
\label{diracres}
BR[(Z \mu^-) \to (Ze^-) \nu_\mu \bar{\nu}_e] &=&
1.19 \times 10^{-6}  \ , \ (\mbox{finite nucleus},
 A=200) \ ,\nonumber \\
BR[(Z \mu^-) \to (Ze^-) \nu_\mu \bar{\nu}_e] &=&
1.19 \times 10^{-6}  \ , \ (\mbox{point nucleus}) \ .
\end{eqnarray}
We note that the effects of the finite size of the nucleus
do influence the shape of the momentum spectrum
$d \Gamma/d|\vec{q}\,|$, as shown in Fig.~4,
but that they leave the total rate
essentially unchanged even for  $Z$ as large as 80.
As the finite size effects are expected to be smaller
for smaller $Z$, we can calculate the rate for $Z=10$ using the
Dirac wavefunctions for a point nucleus.
We get
now, assuming  $\approx$ equal rates for capture and free-electron
decay
\begin{equation}
\label{diracresa}
BR[(Z \mu^-) \to (Ze^-) \nu_\mu \bar{\nu}_e] =
1.25 \times 10^{-9}  \ , \ (\mbox{point nucleus}, Z=10) \ .
\end{equation}

To appreciate the relativistic enhancement due to the Dirac
wavefunctions, we also show the result obtained with the
Schr\"odinger
wavefunctions for a point nucleus:
\begin{eqnarray}
\label{schrres}
BR[(Z \mu^-) \to (Ze^-) \nu_\mu \bar{\nu}_e] &=&
2.32 \times 10^{-8}  \ , \ (Z=80) \ ,  \nonumber \\
BR[(Z \mu^-) \to (Ze^-) \nu_\mu \bar{\nu}_e] &=&
9.32 \times 10^{-10}  \ , \ (Z=10) \ .
\end{eqnarray}
The relativistic enhancement for $Z=80$ is a remarkable factor of 50.

In Fig.~4 the decay distribution $d\Gamma/d|\vec{q}\,|$,
divided by the total width $\Gamma_{tot}$,
is given for $Z=80$.
For illustration we also give
the spectra predicted using
the Schr\"odinger and Dirac wavefunctions for a point nucleus.
Comparing these two curves
shows that relativistic effects not only change
the  overall normalization but cause the spectrum to peak at
higher momenta.  The finite nuclear size causes the shape of the
spectrum as calculated for a point nucleus to narrow.
This can easily be understood by looking at the muonic wavefunction
in Fig.~2.
In Fig.~5 the same distribution is shown for $Z=10$; as might be
expected both relativistic effects and the effect of a finite nuclear
size are small.
Transitions of the form $(Z\mu) \to (Ze) + \nu \bar\nu$
have as their signature a bound state
recoiling with a large momentum the order of $m_\mu$.
The momentum distribution
for heavy atoms ( $Z=80$ )
peaks at about 20 MeV as seen in Fig.~4.

\vspace*{1cm}
{\bf 4.) Conclusions}
\vspace{.8cm}

In this paper we have shown that transitions between
electromagnetic bound states can be calculated reliably
using relativistic
atomic wavefunctions. The rates and also the shapes of the spectra
depend drastically on relativistic corrections;
for heavy muonic atoms these enhance the branching ratio to
a sizeable value of $\sim 10^{-6}$ from $\sim 10^{-8}$
(see Eqs.~(\ref{diracres}) and (\ref{schrres})).
The QED analysis of these alchemy transitions
illustrates some of the physics of the relativistic wavefunctions
that must invariably enter the QCD analysis of the
corresponding exclusive
electroweak decays of hadrons.

Prospects for an experimental test of atomic alchemy are dim but not
hopeless.  While the branching ratio appears to be too small to be
detectable for $(\pi e)$ or for light muonic atoms, it may be possible
to detect it for heavy muonic atoms.
One promising approach is to inject and capture
$\sim 5$~kev muons in a cyclotron trap~\cite{Simons}
containing $\sim 0.1$~bar
of a noble gas, which we will here assume to be neon ($Z=10$).
The muons will come to rest~\cite{Simons} in $\sim 1$~$\mu{\rm s}$
and be captured by a neon atom.  The muons will rapidly cascade to
the $1S$ state, mostly by the ejection of atomic electrons.
In neon the
mean number of electrons in the K~shell
has been measured \cite{Bacher} to
lie between $0.07$ and $0.68$ by the time the muon has fallen
to states with principal quantum number $n = 5$, in agreement with
simple estimates~\cite{Vacant} of 0.25,
and will certainly diminish further
as the muon continues to the $1S$ state.
Total ionization of an atom has been observed
in the cascades of antiprotons in neon,
argon, and krypton~\cite{antipro}.
The electron K shell will therefore be vacant to receive the
electron from atomic alchemy.  Only $0.31$ of the muons in the
$1S$ state of neon will be lost to muon capture~\cite{Muneon}, so most of
the captured muons will be available to decay by atomic alchemy.
At a pressure of $\sim 0.1$~bar, using estimates in \cite{Vacant} for
the velocity of
muonic neon ion ($\sim 10^5\ {\rm cm/s}$) and for the cross section
for electron capture for ${\rm Ne}^{9^+}$ on ${\rm Ne}$
($\sim 3\cdot 10^{-15}\,{\rm cm}^2$), the time taken for a bare neon atom
to capture an electron is the order of $1\ \mu{\rm s}$, so most
of the captured muons will decay before the electron K~shell can refill
and block the atomic alchemy.
It has been suggested~\cite{Bacher} that cyclotron
traps, admittedly with higher gas pressures, can be built
capable of stopping $\sim 10^7\ \mu^-/{\rm s}$, so
a branching ratio as small as $\sim 10^{-9}$ might be accessible.

Even if such a system could be realized
the experimental difficulties would be daunting.  The signature for
atomic alchemy is the appearance of ${\rm {}^{20}{Ne}^{9+}}$
ions with a distribution $dN/dq^2$ extending to a momentum the
order of $m_\mu$.
These ions will be difficult to distinguish from
${\rm {}^{20}F^{9+}}$ ions with a momentum of $m_\mu$
made by internal conversion.  Nor will it be easy to extract the ions
for analysis because in $0.1$ bar of Neon the distance a bare ion can
fly before capturing an electron is only $\sim 1\ {\rm mm}$, and the
gas pressure cannot be lowered without letting the muons in the trap
decay in flight before they can be captured.
However,
measurement of the spectrum $dN/dq^2$ would be valuable;
it
directly reflects the local structure of the
atomic momentum space
wavefunction  and
the beta decay spectrum.
This appears to be one of
the few ways in which one
can directly study the relativistic
tail of bound state wavefunctions in QED.

For induced reactions, such as
$\gamma \, (d^+e^-) \to (p^+e^-) n \, $, one
will observe a nearly monoenergetic final state hydrogen atom
recoiling against an outgoing neutron. In principle
one can use the doppler-shifted radiation from the
outgoing atoms as a precise calibration of the nuclear
scattering kinematics. One could also analyze the spin states
of the outgoing atomic system and its hyperfine spectrum
as a probe of the spin state of the final state nucleus.

\vspace*{1cm}
We would like to thank P. Tru\"ol, R. Engfer,
I.B. Khriplovich, and M. Strikman for
stimulating discussions.

\section*{Figure captions}

\subsection*{Figure 1}
General diagram for a weak alchemy transition
$B_1 \to B_2 + X$
as discussed in
the text. The momenta of the particles are indicated in
parentheses.
\subsection*{Figure 2}
Schr\"odinger and Dirac wavefunctions (multiplied by
$|\vec{k}|$) for a $(Z \mu^-)$ atom with Z=80.
\subsection*{Figure 3}
Schr\"odinger and Dirac wavefunctions (multiplied by
$|\vec{k}|$) for a $(Z e^-)$ atom with Z=80.
\subsection*{Figure 4}
Decay spectrum $d\Gamma/d|\vec{q}\,|$ divided by $\Gamma_{tot}
= 30 \times \Gamma_{e,free}$
(Eq.~(\ref{totalwidth})) for Z=80 (see text).
The full
line is obtained by using Dirac wavefunctions
which take into account the finite size of the nucleus.
The dashed (dashed-dotted) line corresponds
to point-like Dirac (Schr\"odinger) wavefunctions.
The curve based on Schr\"odinger wavefunctions is
multiplied by a factor 50.
\subsection*{Figure 5}
Decay spectrum $d\Gamma/d|\vec{q}\,|$ divided by $\Gamma_{tot}
= 2 \times \Gamma_{e,free}$
(Eq.~(\ref{totalwidth})) for Z=10 (see text).
The full (dashed) line corresponds to
point-like Dirac (Schr\"odinger) wavefunctions.

\subsection*{Appendix: Dirac wavefunctions}

In atomic alchemy $(Z\mu) \rightarrow (Ze)\nu_\mu \bar\nu_\mu$
the electron wavefunction is probed at
momenta the order of $m_\mu$.
Schr\"odinger wave
functions are not appropriate
and the Dirac wavefunctions \cite{BJD1}
must be used.
Recoil corrections due to the
finite nuclear mass are extremely small, so
the Dirac wavefunctions for an electron in the field of an
infinitely heavy nucleus
will describe adequately
the high momentum tail of the electron
(and muon) wavefunction.  At large $Z$ the Bohr radius of
the muon $1S$ state is less than the nuclear radius,
so the effect of the finite size of the nucleus on the muon
wavefunction must obviously be included; the effect of the finite
size is also important on the high-momentum tail of the electron
$1S$ wavefunction.

We sketch
how the momentum-space wavefunctions
$\psi_1(\vec{k})$ and $\psi_2(\vec{k})$
that appear
in the form factors in Eq.~(\ref{formgeneric})
are extracted from
the usual 4-component Dirac wavefunctions in position space.
Because everything can be worked out analytically
for a point nucleus, we describe this case first.
\subsection*{Point nucleus}
 From the literature
(e.g. from \cite{BJD1} on page 55)
we take  the (four component)
ground state wave
function in position space,
$\psi_{n=1,j=1/2,j_z=1/2}(r,\theta,\phi)$.  For brevity we
write this as
$\Phi(\vec{x})$.
Taking the Fourier transform, we get the wavefunction in
momentum space,
\begin{equation}
\label{fourdef}
\tilde{\Phi}(\vec{k}) = \int \, d^3x \, \Phi(\vec{x}) \,
\exp(-i\vec{k}\cdot\vec{x}) \quad ,
\end{equation}
\begin{equation}
\label{fourexpl}
\tilde{\Phi}(\vec{k}) = \left(
\begin{array}{c}
{\displaystyle \hat{f}(k)
\left( \begin{array}{c} 1 \\ 0  \end{array} \right)} \\
\noalign{\smallskip}
{\displaystyle \hat{g}(k) \, \frac{\vec{\sigma}\cdot\vec{k}}{k} \,
\left( \begin{array}{c}  1 \\ 0  \end{array} \right)}
\end{array}
\right)\ , \qquad k=|\vec{k}| \ ,
\end{equation}
with
\begin{eqnarray}
\label{gfun}
\hat{f}(k) &=& \frac{N}{k \left( 1+a^2k^2 \right)^{1+\gamma/2}} \,
(\sin\rho + ak \cos\rho) \\
\label{ffun}
\hat{g}(k) &=& \frac{N m (1-\gamma)}{\gamma k^2 \left( 1+a^2k^2
\right)^{1+\gamma/2}} \, \left[
\left( 1+(1+\gamma) a^2 k^2 \right) \, \sin \rho - \gamma a k
\cos \rho \, \right]
\end{eqnarray}
\begin{equation}
a= \frac{1}{m Z \alpha} \ , \
\gamma = \sqrt{1 - (Z\alpha )^2} \ , \
\rho = \gamma \, \mbox{atan}(ak) \ , \
N = 2^{\gamma+1} \, \Gamma(1+\gamma) \,
\sqrt{\frac{a \pi (1+\gamma)}{\Gamma(1+2\gamma)}} \quad .
\end{equation}
Here $m$ denotes the reduced mass of the system, which we take
to be identical to that of the muon or the electron because we work
to lowest order in $1/M$.
The corresponding energy eigenvalue
is then $E=m \, \gamma$. For $Z=80$ the numerical value of
$\gamma$ is $0.81$.

The Dirac wavefunction for a bound state of course
projects onto plane waves of
both positive and negative energies;
the latter waves correspond to antiparticles.
We therefore
define the relevant wavefunction (to be used in calculating the
form factors) by the projection on positive energy plane waves. Thus
we expand
$\tilde{\Phi}(\vec{k})$ in terms of spinors
$u_{r}(\vec{k})$ and $v_{r}(-\vec{k})$, writing
\begin{equation}
\label{decomp}
\tilde{\Phi}(\vec{k}) = \sum_r \left[
A_r(\vec{k}) \frac{u_r(\vec{k})}{\sqrt{2k^0}} +
B^\star_r(-\vec{k}) \frac{v_r(-\vec{k})}{\sqrt{2k^0}}
\, \right] \  , \quad k^0 = \sqrt{\vec{k}^2 + m^2} \quad .
\end{equation}
If $j_z=1/2$ we get
\begin{eqnarray}
\label{abkoeff}
A_{+1/2}(\vec{k}) &=& \phantom{-}\sqrt{\frac{k^0+m}{2k^0}} \,
\left( \hat{f}(k) +
\frac{k}{k^0+m} \, \hat{g}(k) \, \right) \ ; \quad
A_{-1/2} = 0\ ; \\
B^\star_{+1/2}(-\vec{k}) &=& -\sqrt{\frac{k^0+m}{2k^0}} \,
\frac{ (k^1+i k^2)}{k} \, \left( \frac{k}{k^0+m} \, \hat{f}(k)
  - \hat{g}(k)
\right) \nonumber \\
B^\star_{-1/2}(-\vec{k}) &=& \phantom{-}\sqrt{\frac{k^0+m}{2k^0}} \,
\frac{  k^3}{k} \, \left( \frac{k}{k^0+m} \, \hat{f}(k) -
\hat{g}(k)
\right) \nonumber
\end{eqnarray}
Here
$A_r(\vec{k})$ is the probability amplitude to find an
electron with momentum~$\vec{k}$ and spin~$r$ in the atom, while
$B^\star_r(-\vec{k})$ is the probability amplitude to find a
positron with momentum~ $\vec{k}$ and spin~$r$; the latter amplitude
arises from the creation of $e^+e^-$ pairs on the nucleus.  Because
the wave
function in position space is normalized as
$\int  d^3x \,\Phi^+(\vec{x}\,) \, \Phi(\vec{x}\,)=1$,
the Fourier transform is
automatically normalized as
$\int  d^3k/(2\pi)^3 \, \tilde{\Phi}^+(\vec{k}) \,
\tilde{\Phi}(\vec{k})=1$.  Therefore $A_r(\vec{k})$
and $B^\star_r(-\vec{k})$ are normalized so that
\begin{equation}
\label{normaliz}
\int \, \frac{d^3k}{(2\pi)^3} \, \sum_r \, \left\{
|A_r(\vec{k})|^2 + |B_r(\vec{k})|^2 \right\}
= 1 \quad .
\end{equation}
The integral $\int \, d^3k/(2\pi)^3 \,
\sum_r \, |B_r(\vec{k})|^2 \, $
gives the probability to find a three particle Fock state
$(e^+e^-e^-)$ in the atom. Even for $Z=80$ this fraction
is tiny $(\approx 0.2\%)$,
so we only consider the one-Fock
contribution characterized by $A_r(\vec{k})$. We mention that if
we consider the atom with $j_z=-1/2$, we get
$A_{+1/2}(\vec{k})=0$, and $A_{-1/2}(\vec{k})$ is identical
to $A_{+1/2}(\vec{k})$,
given in Eq.~(\ref{abkoeff})
for $j_z=+1/2$.
The wavefunction denoted in the text
as
$\psi(\vec{k})$
is therefore given in the relativistic case by
\begin{equation}
\label{psirel}
\psi(\vec{k}) = A_{+1/2}(\vec{k})  \quad .
\end{equation}
Because
the coefficients $B_r$ are very small,
the effects of antiparticle production are small, and
we get essentially the same form factor
as we would have got had
we naively computed
the simple overlap of the wavefunctions in position space.

\subsection*{Finite-size nucleus}
As usual (see e.g. \cite{Behrends}),
we model a nucleus of atomic number $A$ as a
homogeneously
charged sphere
of radius $r_0 = 1.3 \, A^{1/3} \, \mbox{fm}$; numerically $r_0 \approx
7\,\mbox{fm}$ for $A = 200$.
Inside the sphere
the potential has the form
$V(r)=-(Z\alpha/r_0) \cdot  (3-r^2/r_0^2)/2$, and
outside we have $V(r)=-Z\alpha/r$.
In the notation of Landau-Lifschitz \cite{LL} we write
the four-component Dirac function $\Phi(\vec{r})$ as
\begin{equation}
\label{fourlandau}
\Phi(\vec{r}\,) = N \, \left(
\begin{array}{c}
{\displaystyle f(r)
\left( \begin{array}{c} 1 \\ 0  \end{array} \right)} \\
\noalign{\smallskip}
{\displaystyle   - i g(r) \, \frac{\vec{\sigma}\cdot\vec{r}}{r} \,
\left( \begin{array}{c}  1 \\ 0  \end{array} \right)}
\end{array}
\right)\ , \quad r=|\vec{r}\,| \ ,
\end{equation}
where the constant~$N$ is chosen such that
$\int \, d^3r \, \Phi^+(\vec{r}\,) \, \Phi(\vec{r}\,) = 1 $.
The radial equations for $f$ and $g$ are
\begin{eqnarray}
\label{radial}
f'(r) - (E+m-V(r)) g(r) &=& 0 \nonumber \\
g'(r) +\frac{2}{r} g(r) + (E-m-V(r)) \, f(r) &=& 0
\end{eqnarray}
which
are solved separately in the two regions $r \le r_0$ and
$r \ge r_0$ for an arbitrary constant
$E < m$. In the outer region
$r \ge r_0$ there is
(up to an overall constant)
exactly one solution $(f,g)$
that is square integrable at
$r=\infty$. It is given by (see Landau-Lifschitz \cite{LL})
\begin{eqnarray}
\label{radialsol}
f(r) &=& \phantom{-}\sqrt{m + E} \, \exp(-\lambda r) \,
r^{\gamma-1} \, (Q_1 + Q_2) \nonumber \\
g(r) &=& - \sqrt{m - E} \, \exp(-\lambda r) \,
r^{\gamma-1} \, (Q_1 - Q_2) \nonumber \\
Q_1 &=& \phantom{-(1 - \frac{Z \alpha m}{\lambda}) \,}
U(\gamma - \frac{Z \alpha E}{\lambda}, 2 \gamma + 1, 2
\lambda r) \nonumber \\
Q_2 &=& -(1 - \frac{Z \alpha m}{\lambda}) \,
U(\gamma + 1 - \frac{Z \alpha E}{\lambda}, 2 \gamma + 1, 2
\lambda r) \nonumber \\
\lambda &=& \sqrt{m^2-E^2} \quad , \quad \gamma =
\sqrt{1- (Z\alpha)^2} \quad .
\end{eqnarray}
The hypergeometric function $U$ is defined and discussed
in detail in chapter~13 of Abramowitz-Stegun \cite{Abramowitz};
it is also related to Whittaker functions which
are numerically accessible in the CERN-library.

In the inner region $(r \le r_0)$
a simplified solution
for the (massless) electron has
been obtained by Khriplovich \cite{Kriplowitch};
however, the substantial
mass of the muon requires a more
complete treatment.
It turns out that the Dirac equation has
(again up to an overall constant)
exactly one
solution $(f,g)$
that is
square integrable at $r=r_0$; in the present case
it has a Taylor series expansion around $r=0$.
While $f$ starts as a constant, $g$
begins with
a term linear in $r$. The coefficients of the power
series expansions of $f$ and $g$ may be
defined recursively.

The inner and outer solutions must satisfy
the matching condition
\begin{equation}
\label{matching}
(f/g)_{r \to r_{0-}} = (f/g)_{r \to r_{0+}}
\end{equation}
in order to be solutions of the complete equation.
This can only hold for certain values of $E$; these are
just the eigenvalues.
For $Z = 80$ and $A = 200$ the lowest eigenvalues are
$E=0.908 m_\mu$ and $E=0.811 m_e$ for the $1S$ states of the
muon and the electron, respectively.

The Fourier transform $\tilde{\Phi}(\vec{k})$ is then defined
and written in terms of
$\hat{f}(k)$ and $\hat{g}(k)$ precisely as in
eqs.~(\ref{fourdef}) and~(\ref{fourexpl}).
Proceeding through the same steps as for a point nucleus,
one finally
gets the wavefunction $\psi(\vec{k})$
(compare with Eq.~(\ref{psirel})) in a numerical form.
This wavefunction
shown for the $(Z \mu)$ and the $(Z e)$ atoms in
Figs.~2 and~3, respectively.

\begin{thebibliography}{12}

\bibitem{Migdal}
    A. B. Migdal,
          ``Qualitative Methods in Quantum
        Theory", Benjamin, 1977,
                (Frontiers in Physics
                     series, 48).

\bibitem{GW}
    C. Greub and D. Wyler, Phys. Lett. {\bf B295} (1992) 293.

\bibitem{aadd}
    See ``Weak Interaction Physics'',
    C. Rubbia, in ``High Energy Physics,'' E.H.S. Burhop, ed.,
    Academic Press, New York, 1969, Table VI.

\bibitem{Simons}
     L.~M.~Simons, in
     ``Fundamental Symmetries'', P.~Bloch, ed., p.~89,
     Plenum Press, New York, 1987.  See also \cite{Bacher}.

\bibitem{Bacher}
      R.~Bacher, P.~Bl\"um, D.~Gotta,
      K.~Heitlinger, and M.~Schneider,
      Phys. Rev. A{\bf 39},~1610~(1989).

\bibitem{Vacant}
      R.~Bacher, D.~Gotta, L.~M.~Simons, J.~Missimer, and
      N.~M.~Mukhopadhyay,
      Phys. Rev. Lett.~{\bf 54},~2087~(1985).

\bibitem{antipro}
      R.~Bacher, P.~Bl\"um, D.~Gotta, K.~Heitlinger, M.~Schneider,
      J.~Missimer, L.~M.~Simons, and K.~Elsener,
      Phys. Rev. A{\bf 38},~4395~(1988).

\bibitem{Muneon}
      J.~L.~Rosen,
      E.~W.~Anderson, E.~J.~Bleser, L.~M.~Lederman, S.~L.~Meyer,
      J.~E.~Rothberg, and I.~T.~Wang,
      Phys. Rev. {\bf 132},~2691~(1963).

\bibitem{BJD1}
    J. D. Bjorken and S. D. Drell, ``Relativistic
    Quantum Mechanics", International Series in
    Pure and Applied  Physics, McGraw-Hill Company, New York.


\bibitem{Behrends}
     H. Behrends and W. B\"uhring, ``Electron Radial Wave
     Functions and Nuclear Beta-decay",
     International Series of Monographs on Physics, Vol. 67,
     Clarendon Press, Oxford 1982.

\bibitem{LL}
      L. D. Landau, E. M. Lifschitz,
     ``Course of Theoretical Physics, Volume 4:
      Quantum Electrodynamics" , Pergamon Press, Oxford.

\bibitem{Abramowitz}
      M. Abramowitz and A. Stegun, ``Handbook of Mathematical Functions",
       Dover Publications, Inc., New York.

\bibitem{Kriplowitch}
     I.B. Khriplovich, ``Parity Nonconservation in atomic
     Phenomena", Gordon and Breach, 1991.

\end{thebibliography}
\end{document}